\documentclass[12pt]{iopart}
\usepackage{graphicx}
\newcommand{\ket}[1]{| #1 \rangle}

\newcommand{\be}{\begin{equation}}
\newcommand{\ee}{\end{equation}}
\newcommand{\bea}{\begin{eqnarray}}
\newcommand{\eea}{\end{eqnarray}}
\newcommand{\beas}{\begin{eqnarray*}}
\newcommand{\eeas}{\end{eqnarray*}}
\newcommand{\bes}{\begin{equation*}}
\newcommand{\ees}{\end{equation*}}

\begin{document}

\title{All-optical switching of photonic entanglement}

\author{Matthew A. Hall, Joseph B. Altepeter, and Prem Kumar}
\address{Center for Photonic Communication and Computing, EECS Department\\
Northwestern University, 2145 Sheridan Road, Evanston, IL
60208-3118}
%\author{Joseph B. Altepeter}
%\address{Center for Photonic Communication and Computing, EECS Department\\
%Northwestern University, 2145 Sheridan Road, Evanston, IL
%60208-3118}
%\author{Prem Kumar}
%\address{Center for Photonic Communication and Computing, EECS Department\\
%Northwestern University, 2145 Sheridan Road, Evanston, IL
%60208-3118}

\begin{abstract}
Future quantum optical networks will require the ability to route
entangled photons at high speeds, with minimal loss and added in-band
noise, and---most importantly---without disturbing the photons' quantum state.
Here we present an all-optical switch which fulfills these requirements and
characterize its performance at the single photon level. It exhibits a
200-ps switching window, 120:1 contrast, 1.5-dB loss, and induces
no measurable degradation in the switched photons' entangled-state fidelity
($< 0.002$). As a proof-of-principle demonstration of its capability, we use
the switch to demultiplex a single quantum channel from a dual-channel,
time-division-multiplexed entangled photon stream. Furthermore, because
this type of switch couples the temporal and spatial degrees of freedom, it
provides an important new tool with which to encode multiple-qubit quantum
states on a single photon.
%To deploy and operate a quantum network which utilizes existing
%telecommunications infrastructure, it is necessary to be able to 
%route entangled photons at high speeds, with
%minimal loss and signal-band noise, and---most importantly---without disturbing
%the photons' quantum state. 
%Here we present a switch which fulfills these requirements and characterize its
%performance at the single photon level; it exhibits a 200-ps switching
%window, a 120:1 contrast ratio, 1.5 dB loss, and induces no
%measurable degradation in the switched photons' entangled-state fidelity ($< 0.002$).  
%Furthermore, because this type of switch couples the temporal and spatial degrees
%of freedom, it provides an important new tool with which
%to encode multiple-qubit states in a single photon.
%As a proof-of-principle demonstration of this capability, 
%we demultiplex a single quantum channel from
%a dual-channel, time-division-multiplexed entangled photon stream,
%effectively performing a controlled-bit-flip on a two-qubit
%subspace of a five-qubit, two-photon state.
\end{abstract}

\maketitle

\section{Introduction}

Switching technologies enable networked rather than point-to-point
communications.  Next-generation photonic quantum networks will
require switches that operate with low loss, low signal-band noise,
and \emph{without} disturbing the transmitted photons' spatial,
temporal, or polarization degrees of freedom \cite{mike_and_ike}.  Additionally,
the switch's operational wavelength must be compatible with
a low-loss, non-dispersive transmission medium, such as 
standard optical fiber's 1.3-$\mu$m zero-dispersion band \cite{nweke, oband}.
Unfortunately, no previously demonstrated technology
\cite{previous-first}--\cite{nolm2} is capable of
\emph{simultaneously} satisfying each of the above requirements:
waveguide electro-optic modulators (EOMs) \cite{eospace} and 
resonators \cite{waveguide_res1, waveguide_res2}
can operate at very high speeds (10 GHz) but completely destroy any
quantum information stored in the polarization degree of freedom;
micro-electromechanical switches \cite{mems1, mems2} do not degrade the photon's
quantum state, but operate at very low speeds ($<=250$ kHz);
polarization-independent EOMs \cite{eospace} 
%and acousto-optic devices \cite{acousto-optic} 
can operate at moderate speeds ($\sim$100 MHz) but with
relatively high loss; and finally, traditional 1550-nm devices
based on nonlinear-optical fiber loops \cite{nolm1, nolm2} generate unacceptably high
levels of Raman-induced noise photons ($> 1$ in-band noise photon
per 100-ps switching window \cite{cband_noise}).

Although the requirements for ultrafast entangled-photon switching
are collectively daunting, they describe a device that is capable
of selectively coupling the spatial and temporal degrees of
photonic quantum information. In other words, a device that can
encode multiple-qubit quantum states onto a single photon, 
enabling quantum communication protocols that exploit
high-dimensional spatio-temporal encodings.  In this paper
we describe the construction and characterization of
an all-optical switch which meets each of the aforementioned
requirements.
Moreover, this switch design is scalable: by its extension
one can create devices that are capable of coupling
many temporal qubits and many spatial qubits.  As a proof-of-principle demonstration of
this capability, we utilize the
switch to perform a controlled-bit-flip operation on a two-qubit subspace
of a two-photon, five-qubit system, where a temporally encoded qubit is
used as the control and a spatially encoded qubit is used as the target.
This operation is used to demultiplex a single
quantum channel from a dual-channel, time-division-multiplexed
entangled photon stream encoded into the larger five-qubit space.  

\subsection{Comparison with existing switching technologies}

\begin{figure}
\centering
\includegraphics[width=4.5in]{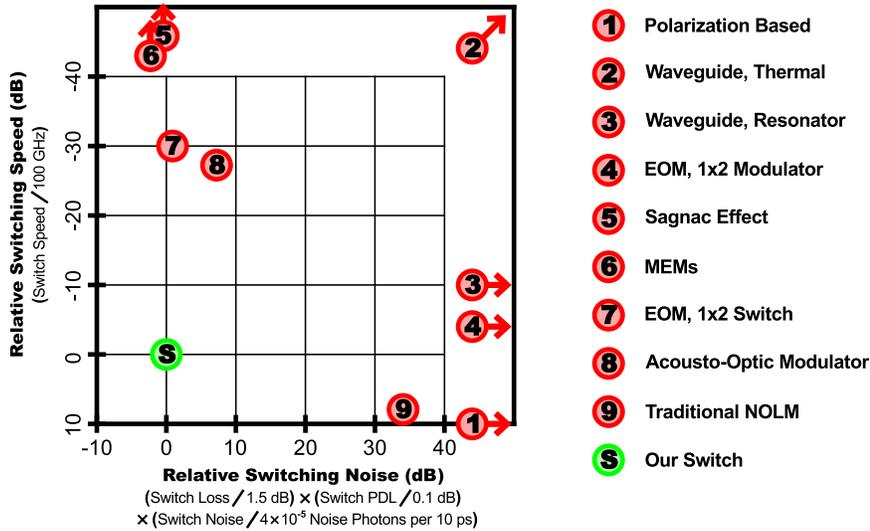}
\vspace{-1mm}
\caption{
Relative performance of switching technologies
in terms of speed (defined in dB
units as the ratio of the two speeds) and aggregate noise (defined
in dB units as ``relative loss for each polarization'' + ``relative production rate of in-band
noise photons'').
} \label{figure::orders}
\vspace{-3mm}
\end{figure}

The aggregate performance of this switch (in terms of
loss, speed, and in-band noise) exceeds that of all available
alternatives \cite{previous-first}--\cite{nolm2} by orders of magnitude.
Fig. \ref{figure::orders} shows the relative performance of our switch
compared to each of these alternatives in terms of speed (defined in dB
units as the ratio of the two speeds) and aggregate noise (defined
in dB units as ``relative loss for each polarization'' + ``relative production rate of in-band
noise photons'').
This figure compares our switch to
ten telecom-band switching technologies:
micro-electromechanical (MEM) switches (very low speed) \cite{mems1}, 
Sagnac effect switching (very low speed)
\cite{previous-last},
polarization interferometer (low speed, low bandwidth,
requires phase stabilization) \cite{eospace}, acousto-optic modulation (low
speed, low bandwidth, high loss) \cite{sintec}, various polarization-based
switches (fundamentally single-polarization, destroying all
polarization-encoded quantum information) \cite{polarization_based}, waveguide resonators
(high loss, low bandwidth, single polarization) \cite{waveguide_res1,
waveguide_res2}, thermal waveguides
(low speed, high loss, single polarization) \cite{waveguide_thermal},
electro-optical modulators or EOMs
(low bandwidth, single polarization) \cite{eospace}, and traditional
nonlinear optical loop mirror (NOLM) switches
(high signal band noise) \cite{nolm1, nolm2}.
Considering only those categories
which are essential for entangled-photon switching (speed, loss,
polarization-dependence,
and signal-band noise), our switching
design exceeds the performance of \emph{every} alternative---including
traditional NOLM-based switches---by at least 33 dB (combined
over all four categories).

\begin{figure}
\centering
\includegraphics[width=4.7in]{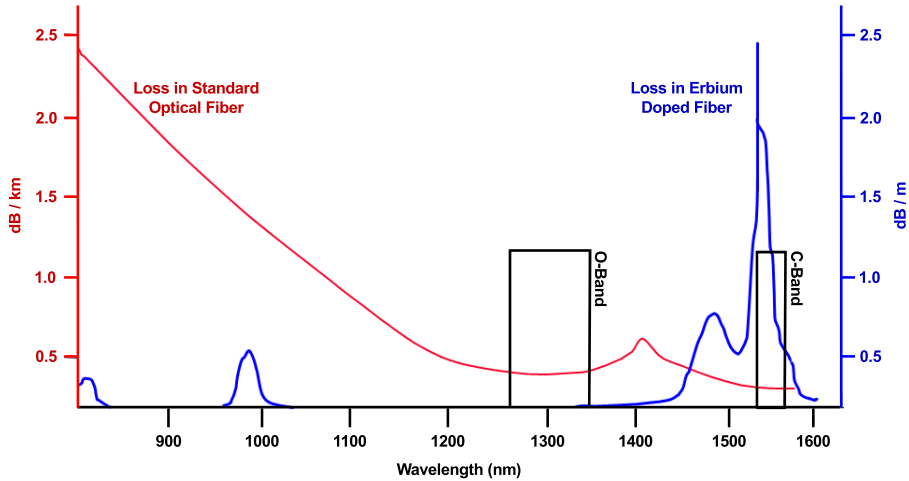}
\vspace{-1mm}
\caption{
%The absorption bands of both standard optical fiber and Erbium-doped fiber,
%commonly used in C-band optical amplifiers.  Notice that the 1.3-$\mu$m
%O-band suffers from minimal loss in both types of fiber.
Absorption band of standard optical fiber (red, left ordinate) and that of
Erbium-doped fiber commonly used in C-band optical amplifiers (blue, right
ordinate). Notice that the 1.3-$\mu$m O-band provides minimal loss in both
types of fiber.
} \label{figure::o-band}
\vspace{-3mm}
\end{figure}

Our switch is designed to operate on quantum signals (e.g., single photons)
in the 1.3-$\mu$m O-band (as opposed to the 1.5-$\mu$m C-band).  This
choice was made for several reasons. 
Foremost, the dispersion minimum for standard optical fiber falls in the
O-band, which avoids dispersive decoherence of the high-speed quantum
signals passing through the switch. Moreover, the O-band is not only close
to the loss minimum in standard optical fiber, signals in this band also
traverse without loss through Erbium-doped fiber that is common in C-band
optical amplifiers (see Fig. \ref{figure::o-band}).
%The O-band
%not only minimizes loss in standard optical fiber, but also
%in the type of Erbium-doped fiber common in C-band optical amplifiers (see
%Fig. \ref{figure::o-band}).  Moreover, the O-band contains the dispersion
%minimum for standard optical fiber, avoiding dispersive decoherence for
%quantum signals.  
Finally, because it is possible to transmit O-band
entangled photons through an active C-band optical amplifier \cite{oband}, it is
conceivable that O-band quantum signals and C-band classical signals could
co-exist in the standard telecommunications infrastructure.

Note that the traditional NOLM-based C-band devices are unsuitible for single-photon
switching for two reasons:  Firstly, and most importantly, 
such switches generate very high levels of Raman-induced background
photons at signal wavelenths \cite{cband_noise}.  
These noise photons would swamp 
any single-photon signals, effectively ``washing out'' any two-photon quantum correlations.
Secondly, most NOLM-based device designs utilize pump pulses which are
group-velocity matched to the signals being switched.  While this increases
the interaction time, the nonlinear character of
the cross-phase modulation (XPM) process, on which such devices rely, 
limits the switching contrast in this type of operation.
Because the pump pulse can not be made perfectly square-shaped,
the center
of the signal pulse receives a stronger nonlinear phase shift than the pulse
wings, making it effectively impossible to choose a single pump power which
maximizes switching contrast over the entire signal pulse.

\section{Switch Design}

\begin{figure}
\centering
\includegraphics[width=5in]{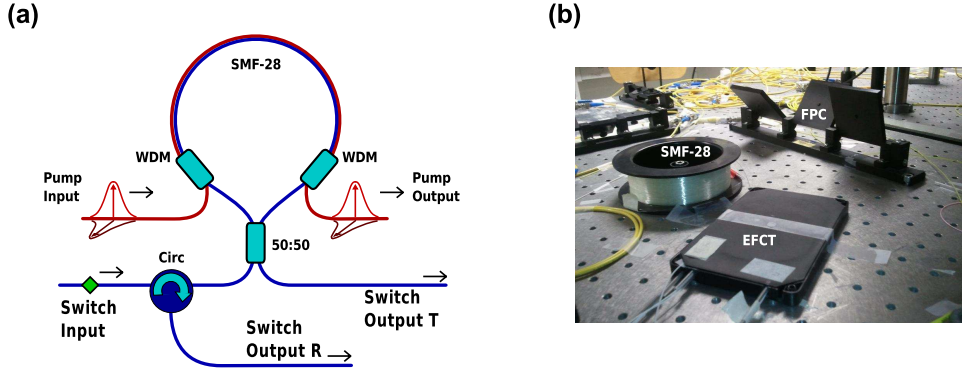}
\vspace{-1mm}
\caption{
(a) The single-photon switch. The length of the intra-loop SMF-28 fiber is directly
proportional to the switching window. An $L$ = 100-m loop results
in a $\approx$200-ps switching window. 
(b) A photograph of the prototype switch.  The enclosed fiber component
tray (EFCT) contains the
50/50 coupler, circulator, and wavelength division multiplexers.
FPC, fiber polarization controller.
} \label{figure::switch_desc}
\vspace{-3mm}
\end{figure}

In order to simultaneously achieve low loss and ultrafast switching, we
utilize an all-optical, fiber-based design in which bright 1550-nm pump
(C-band)
pulses control the trajectory of 1310-nm (O-band) single-photon signals
(see Fig.
\ref{figure::switch_desc}(a)).  Physically, this switch exploits polarization-insensitive cross-phase
modulation \cite{xpm} in a nonlinear-optical loop mirror (NOLM)
\cite{dual-lambda}, the
reflectivity of which is determined by the phase difference between the
clockwise and counter-clockwise propagating paths in a fiber Sagnac
interferometer (the ``loop'') \cite{loop-mirror}.  To actively control the state of
this switch, we initially configure an intra-loop fiber polarization controller 
such that the loop \emph{reflects} all incoming photons.  Multiplexing
a strong 1550-nm pump pulse into the clockwise or counter-clockwise loop
path then creates an XPM-induced phase shift on the respective clockwise or
counter-clockwise signal amplitude, with a $\pi$ phase shift causing the switch
to \emph{transmit} all incoming photons.  

\begin{figure}
\centering
\includegraphics[width=4.5in]{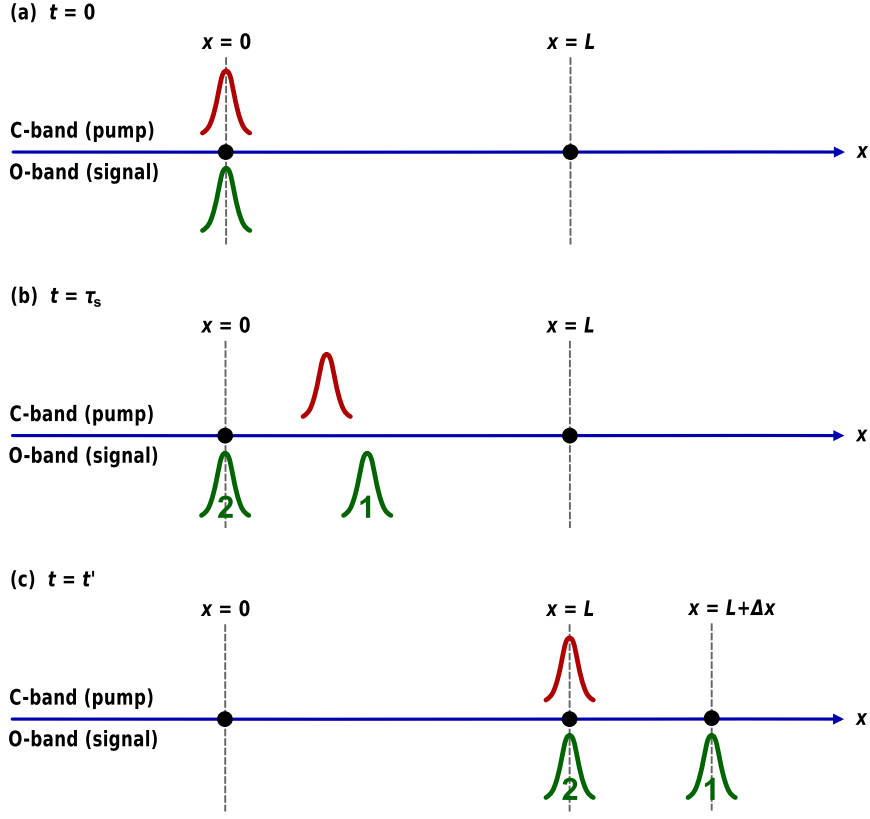}
\vspace{-1mm}
\caption{
Pictorial diagram of a pump pulse and two signal pulses traveling clockwise
through the switching loop for an example switching operation.  
Here the loop is depicted as a single straight
path of length $L$, and the signal and pump pulses are assumed to have
negligible temporal widths.  
(a) For this example, consider that the pump pulse
(C-band / 1550 nm) and the first
signal pulse (O-band / 1310 nm) enter the switching region (via a wavelength division
multiplexer) at $t = 0, x = 0$.  
(b) At time $\tau_s$, the second signal pulse enters the switching region.
(c) By the time the pump pulse exits the switching region at $t = t^\prime,
x = L$ (via another
wavelength division multiplexer), dispersion in the fiber causes
the first signal pulse to travel some additional distance $\Delta x$ and
the second signal pulse to exactly catch up to the pump.  
} \label{figure::pump_signal_timing}
\vspace{-3mm}
\end{figure}

The magnitude of this phase shift is proportional to the instantaneous
intensity of the pump multiplied by the interaction time.
The length of the switching loop, $L$, the group-velocity
difference between the signal and pump pulses, $\Delta v \equiv v_s - v_p$, and the temporal
profile of the pump pulse itself all affect the final temporal profile of
the switching window (which as a function of time determines the
probability that a signal photon will be transmitted through the switch).
The group-velocity difference, in particular, is crucial for high contrast
switching.  Consider the example in Fig.
\ref{figure::pump_signal_timing}.

In this diagram, the switching loop is depicted as a single straight path
of length $L$, where the pump pulse is multiplexed into the switching fiber
at $x=0, t=0$ and multiplexed out of the switching fiber at 
$x=L, t=t^\prime$.  Because in general a $\chi^{(3)}$ medium will be
dispersive, we can expect there to be a non-zero difference
between the pump pulse's group velocity $v_p$ and the signal pulse's group velocity
$v_s$.  In standard fiber, 1310-nm O-band
pulses travel faster than 1550-nm C-band pulses, causing 
an O-band pulse entering the switching region at the same time
as the pump pulse to travel an additional distance $\Delta x$ by
the time the pump exits the switching fiber at $x=L$.  
Notice that in this example the first signal pulse immediately outpaces
the pump pulse, implying that a cross-Kerr phase shift will only be applied
to the signal near $x=0$.  A second signal pulse which enters \emph{after}
the pump pulse at a
time $\tau_s$ will catch the pump at 
$x=L, t=t^\prime$.   A cross-Kerr phase shift will be applied to this
second pulse near $x=L$.  Note that the temporal extent of the
\emph{switching window} is not dependent on any specific signal pulses, or
even on the existence of signal pulses.  In the example above any O-band
signal pulse
which enters the fiber between $t=0$ and $t=\tau_s$ will be transmitted and
any which enters before or after this window will be reflected.

To calculate the walkoff time $\tau_s$ from the switching fiber parameters, first note that:
\begin{equation}
    t^\prime = \frac{L}{v_p} = \frac{L + \Delta x}{v_s}
\end{equation}
and
\begin{equation}
    \tau_s = \frac{\Delta x}{v_s}.
\end{equation}
By combining these two equations, we can derive an expression for $\tau_s$
in terms of $L$, $v_s$, and $v_p$:
\begin{equation}
    \frac{L}{v_p} - \frac{L}{v_s} = \frac{\Delta x}{v_s} = \tau_s.
\end{equation}
If the pump pulse is a delta function in time, the switching window is a
square wave of temporal width $\tau_s$ and a height proportional to the
pump intensity and inversely proportional to $\Delta v$ (i.e., proportional
to the nonlinear interaction time).

\begin{figure}
\centering
\includegraphics[width=3.25in]{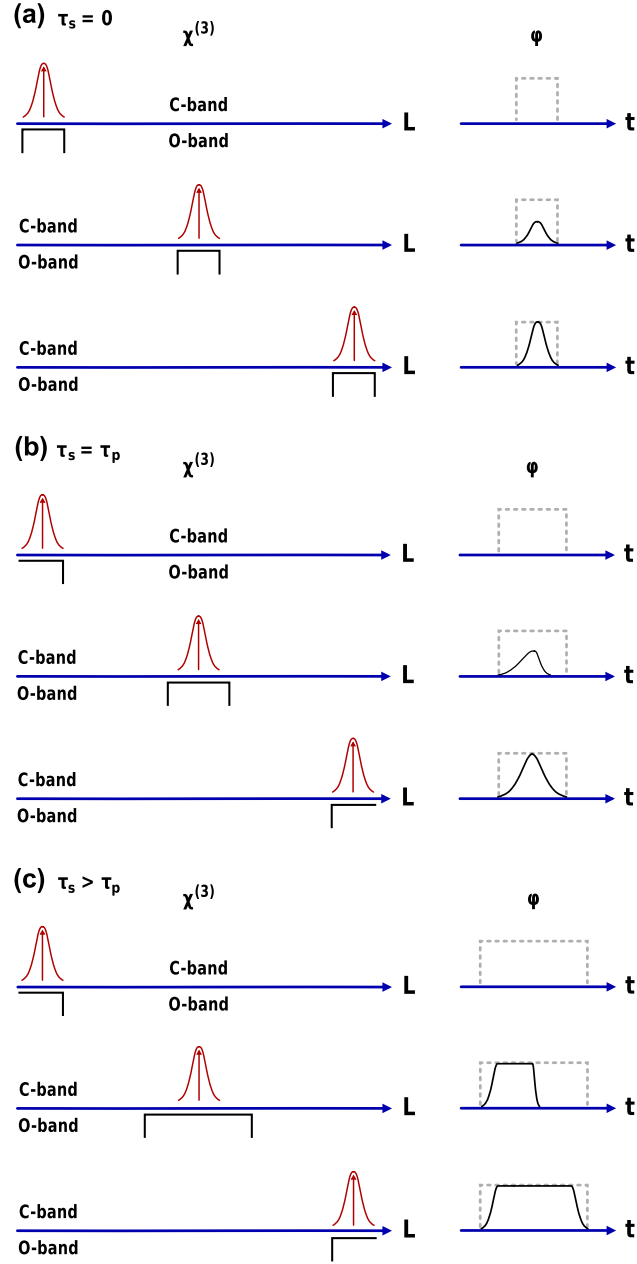}
\vspace{-1mm}
\caption{
Pictorial diagrams of cross-phase accumulation, $\phi$, in the switch as the pump
(control) pulse propagates through the switching fiber for various values
of $\tau_s$.  The left side shows the pump pulse relative
to the target O-band switching window at various points in the switching
fiber.  The right side shows the accumulated cross-phase profile over the
target switching window, indicated by a dotted rectangle of width $\tau_s
+ \tau_p$.  The rectangle width indicates the
target window's temporal width while its height 
indicates the phase necessary to attain full switching.
(a) $\tau_s = 0$.  The cross-phase accumulates linearly with the length of the fiber,
but its shape matches that of the pump pulse's temporal profile.
(b) $\tau_s = \tau_p$.  The
relative delay accumulated between the pump and the signal pulses is equal
to the temporal extent of the pump pulse, leading to a nonuniform cross-phase shift.
(c) $\tau_s > \tau_p$. Some portion of the signal window walks completely
through the pump pulse, leading to a region of uniform (flat) cross-phase (i.e.,
uniform switching) with width $\tau_s - \tau_p$.
} \label{figure::delta-v}
\vspace{-3mm}
\end{figure}

For real pump pulses, the switch can operate
in three general regimes, depending on the pump pulse width $\tau_p$ and
the walkoff time $\tau_s$, which is in turn
determined by the dispersion properties of the switching medium.  If
$\tau_s = 0$, which will be the case when there is no group-velocity difference
between the signal and pump pulses, then the two pulses propagate in
lock-step through the entire
switching medium until the pump is removed via a wavelength-division
multiplexer (WDM).  In principle, this setting should maximize the interaction time,
making it easier to achieve the required cross-phase shift of $\pi$, while at the
same time
minimizing the switching window; in fact, for this case the temporal extent of the switching
window will be identical to the temporal extent of the pump pulse.  In
practice, this configuration has a serious disadvantage, however,
as real pump pulses do not have square-shaped temporal profiles but are in
general Gaussian or bell-shaped.
Such pulses would impart a nonuniform cross-phase shift to the signal pulse, as shown
in Fig.
\ref{figure::delta-v}(a).  As $\tau_s$ increases, the temporal extent of
the switching window
would increase along with the pump intensity necessary to achieve a 
cross-phase shift of $\pi$.  Eventually it approaches a critical point at $\tau_s =
\tau_p$
where the relative delay accumulated between the pump and the signal pulses
(inside the switching fiber) is exactly equal to the temporal extent of the
pump pulse.  At this point the temporal extent of the switching
window is $2\tau_s = 2\tau_p$, and the cross-phase profile is
still highly nonlinear.  This case is shown in Fig. \ref{figure::delta-v}(b).
If $\tau_s$ is increased to $\tau_s > \tau_p$, 
at least some portion of the signal will have
time to completely ``walk through'' the pump pulse, creating a window where the
cross-phase is uniform regardless of the exact shape of the pump
pulse---see Fig. \ref{figure::delta-v}(c).  As $\tau_s$ increases further, this
flat cross-phase region will continue to increase.  In all cases, the final
cross-phase
profile is equal to the convolution of the pump pulse's temporal profile
with a square wave whose width is equal to $\tau_s$.  For standard single
mode fiber and for the wavelengths considered, $\tau_s \approx L \times 1.7$~ps/m.

%Our switching design is immune to each of these two fundamental problems;
%the large spectral detuning ($\approx$35 THz) between the 1550-nm pump pulses and
%the 1310-nm single-photon pulses allows for noiseless, high contrast
%switching operation.  Firstly, the anti-Stokes detuning prevents

The ability to generate a flat cross-phase profile from pump/signal walkoff
is a key reason why this switch is suitable for single-photon switching
while traditional C-band devices are not.  Traditional C-band devices
utilize non-square-wave pump pulses which are group-velocity matched to the
signals being switched (the case shown in Fig. \ref{figure::delta-v}(a)),
which makes it impossible to choose a single pump power which maximizes
switching contrast over the entire signal pulse.  In contrast, the large
group-velocity difference between the signal and pump wavelengths
(determined by the loop medium) allows
our switch to operate in a regime
where the pump pulse walks completely through the signal's temporal mode,
providing the type of uniform phase shift which is essential for
high-contrast switching operation. The effective phase shift is therefore
determined by the total energy in a single pump pulse, regardless of that
pulse's temporal profile.  
The switching \emph{window}, $\tau$, is in turn
determined by the \emph{length of the
fiber} between the WDMs, $L$, multiplied by the speed at which the signal sweeps through the
pump (for standard single-mode fiber, 1310-nm pulses have a higher
group velocity than
1550-nm pulses).
For our case, $\tau = \tau_s + \tau_p \approx \tau_s = L \times 1.7$ ps/m.  
The turn-on time of this switching window is
set by the temporal extent of the pump pulses (i.e., the time it
takes for the pump to physically enter and leave the fiber loop).  

As XPM is inherently polarization
dependent, and polarization is often used to encode quantum information, it
is important that the pump pulse itself be effectively unpolarized.  
We accomplish this by temporally overlapping two orthogonally
polarized pump pulses, each with a slightly different wavelength
\cite{xpm} (1545 nm and 1555 nm).  In the
switch described here, the dual wavelength pumps have a 1-nm bandwidth for
a transform-limited turn-on time of 5 ps.  If we were to instead use a 5-nm
bandwidth pump,
for example, the turn-on time could be as short as 1 ps.
For such pump pulses, the operation of our switch would approach the ideal,
flat phase-profile regime discussed above, even for fiber lengths as short
as 5 m that would lead to switching windows as short as 10 ps.

Finally, although the large group-velocity difference between 1310-nm and
1550-nm wavelengths in
standard single-mode fiber (and the resulting flat switching profile) is a key
advantage of this switch relative to standard C-band designs, it is not the
most important advantage.  The most important advantage our switch has over
traditional designs is the large
anti-Stokes detuning  ($\approx$35 THz) between the 1550-nm pump pulses
and the 1310-nm single-photon pulses, which dramatically reduces
contamination of the quantum channels by spontaneous Raman scattering of
the pump (to a level of $\approx 2 \times 10^{-7}$ background photons per
ps of the signal pulse).

\section{Switch Construction}

The basic components necessary to construct a Sagnac-loop based single-photon switch
(see Fig. \ref{figure::switch_desc}) 
include a 50/50 fiber-coupler (the entrance to the Sagnac interferometer),
a pair of wavelength-division multiplexers (to add and drop the 1550-nm
pump pulses), a fiber polarization controller (to configure the switch to
passively reflect input light), a switching fiber 
%(with a length $L$ equal to 1m/2ps of the desired switching window)
(with a length $L$ in meters equal to one-half of the desired switching
window in ps),
and a circulator (to redirect the reflected
light).
In addition, two key experimental technologies are
required to operate and
characterize this type of switch: a short-pulse 
dual-wavelength 1550-nm pump and a source of 1310-nm entangled
photons.

\subsection{Preparation of dual-wavelength pump pulses}

%To create the dual-wavelength pump, two 5-ps duration pulses (1545-nm and 1555-nm
%wavelengths) are spectrally carved with diffraction gratings from the output of a
%50-MHz repetition rate mode-locked fiber laser (IMRA Femtolite Ultra, Model
%BX-60), which are then multiplexed using a polarization beam combiner (PBC). 
%The power necessary to produce a $\pi$
%phase shift is obtained by amplifying the multiplexed pulses with 
%a cascade of erbium-doped fiber amplifiers
%(EDFAs). A long-pass filter with a 1543-nm edge is used after each EDFA to
%ensure that the optical gain is confined to the pump pulses and that no
%contaminating O-band photons are introduced by the pump preparation process.
%Fig. \ref{figure::dual_color_pump} shows an experimental schematic of the
%dual-color pump preparation apparatus.

\begin{figure}
\centering
\includegraphics[width=5.5in]{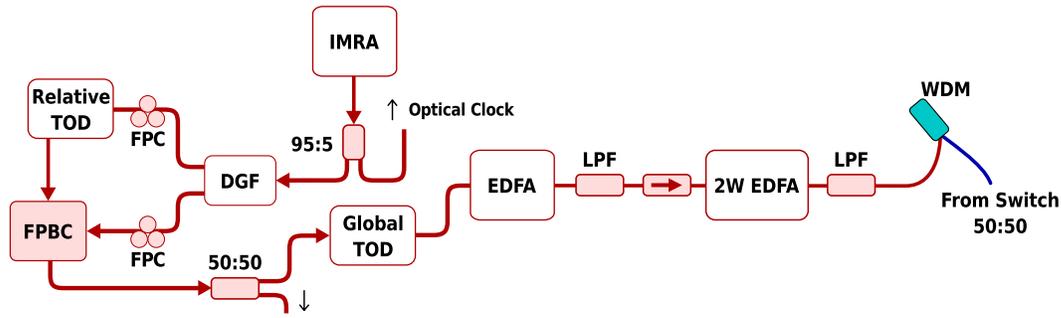}
\vspace{-1mm}
\caption{
Experimental schematic of the dual-color pump preparation process.
IMRA, femtosecond-pulse laser; DGF, double-grating filter; FPC, fiber
polarization controller; TOD, tunable optical delay; FPBC, fiberized
polarization beam combiner; EDFA, Erbium-doped-fiber amplifier; LPF,
long-pass filter; WDM, wave-division multiplexer; and the red horizontal
arrow denotes an optical isolator (a unidirectional optical device).
} \label{figure::dual_color_pump}
\vspace{-0mm}
\end{figure}

Fig. \ref{figure::dual_color_pump} shows an experimental schematic of the dual-wavelength pump
preparation apparatus. Two streams of 5-ps duration pulses (1545-nm and
1555-nm wavelengths) are spectrally carved with a diffraction-grating
filter (DGF) from the output of a 50-MHz rate mode-locked fiber laser (IMRA
Femtolite Ultra, Model BX-60), which are then multiplexed using a fiberized
polarization beam combiner (FPBC). A tunable optical delay (TOD) is
introduced in the path of one wavelength in order to adjust the relative
timing between the two pulse streams. Another TOD is introduced after the
FPBC to adjust the arrival time of the dual-wavelength pulses at the WDM in
the Sagnac loop. The power necessary to produce a $\pi$ phase shift is
obtained by amplifying the multiplexed pulses with a cascade of
erbium-doped fiber amplifiers (EDFAs). A long-pass filter (LPF) with a
1543-nm edge is used after each EDFA to ensure that the optical gain is
confined to the pump pulses and that no contaminating O-band photons are
introduced by the pump preparation process. 

\subsection{Entangled photon source and measurement apparatus}

\begin{figure}
\centering
\includegraphics[width=3.5in]{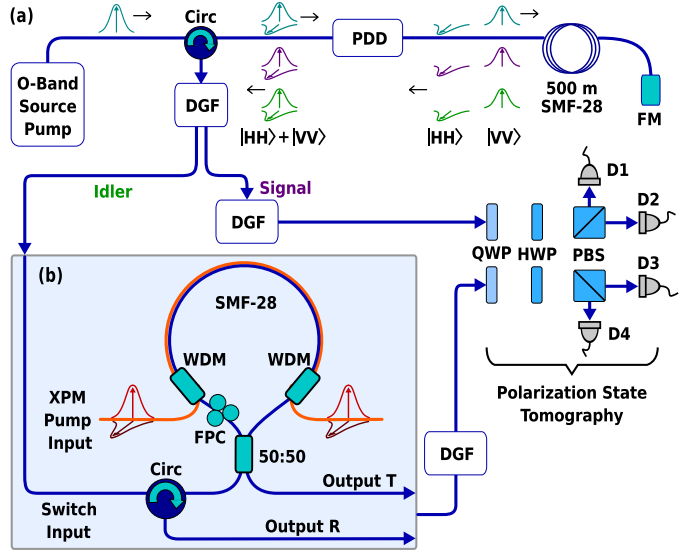}
\vspace{-1mm}
\caption{
(a) Entangled
photon-pair source and test apparatus for ultrafast switching. Nondegenerate
entangled photon pairs ($\lambda_\mathrm{signal} = 1303.5$ nm and
$\lambda_\mathrm{idler} = 1306.5$ nm) are
generated in 500 m of standard single-mode fiber (SMF-28). Signal and idler
photons are separated using a double-pass grating filter (DGF),
with idler
photons then sent to the single-photon switch. Both signal and idler
photons are eventually subjected to polarization-basis tomography. 
Circ: circulator, FM: Faraday mirror, FPC: fiber polarization controller, HWP:
half-wave plate, PBS: polarizing beam-splitter, PDD: polarization dependent
delay, QWP: quarter-wave plate, WDM: wavelength-division multiplexer. (b)
The single-photon switch. The length $L$ of the intra-loop SMF-28 is directly
proportional to the switching window. An $L$ = 100-m loop results
in a $\approx$200-ps switching window. 
%(c) Reconstructed density matrix of the
%unswitched state with the 500-m switch (fidelity to a maximally entangled
%state, $F = 99.5\% \pm 0.2\%$; tangle, $T = 0.982 \pm 0.005$; linear
%entropy, $S_L = 0.0005 \pm 0.003$). 
%(d) Reconstructed density matrix of the switched state
%from the 500-m switch ($F = 99.2\% \pm 0.2\%$, $T = 0.967 \pm 0.007$, 
%$S_L = 0.01 \pm 0.005$). 
%(e) Reconstructed density matrix of the unswitched state for 
%$L$ = 100-m (fidelity to a maximally entangled state, $F = 99.5\% \pm
%0.2\%$, $T = 0.981 \pm 0.01$, $S_L = 0.001
%\pm 0.004$).  
%(f) Density matrix of the switched state for $L$ = 100 m 
%$(F = 99.4\% \pm 0.4\%)$, $T = 0.978 \pm 0.01$, $S_L = 0.005 \pm 0.01$.
} \label{figure::source-switch}
\vspace{-3mm}
\end{figure}

\begin{figure}
\centering
\vspace{3mm}
\includegraphics[width=5in]{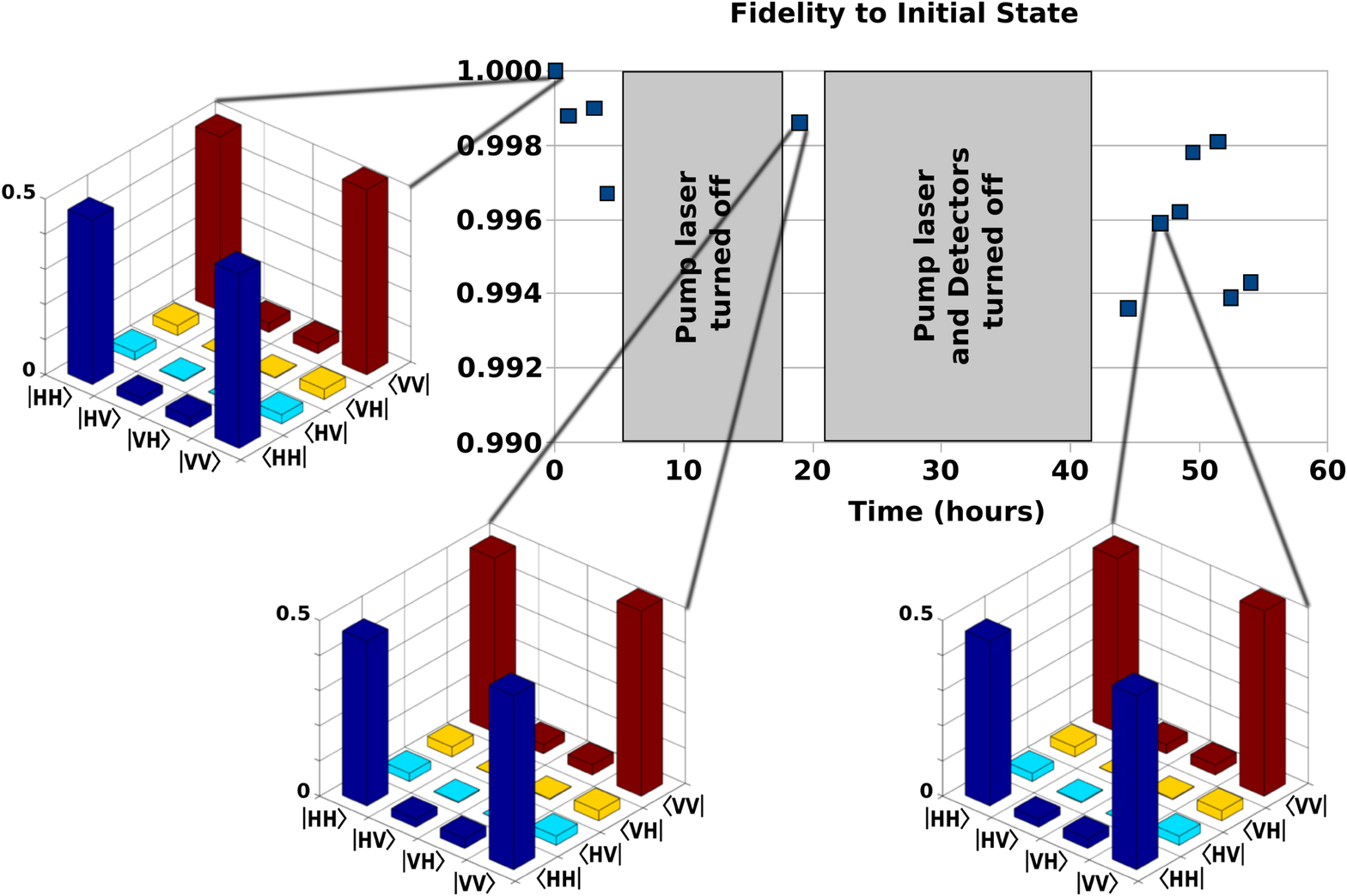}
\vspace{-1mm}
\caption{
Source fidelity to a single maximally entangled state as a function of time.
The fidelities were obtained via tomographic reconstruction of the density
matrices of the entangled photons emitted by the source (a few
representative density matrices are plotted as insets). Tomographic
measurements were made at somewhat periodic intervals over the course of
two and a half days. No realignment was performed during this period even after
power cycling the pump laser (gray boxes indicate periods where the source
was completely powered down).
} \label{figure::source_stability}
\vspace{-3mm}
\end{figure}

The IMRA laser also provides an electrical clock signal for a 1310-nm entangled photon
source and an array of four single-photon detectors.  The entangled photon
source, 
shown in Fig.
\ref{figure::source-switch}(a), utilizes spontaneous
four-wave-mixing in standard single-mode fiber to produce pairs of
polarization-entangled photons from
100-ps wide, 50-MHz repetition-rate pump pulses at 1305 nm.  
Although the photon-pair source is described in detail in 
\cite{oband}, it is noteworthy that this type of source design is extremely
robust against environmental perturbations, and is capable of producing an
identical, nearly maximally entangled state without realignment for a
period of several days.  The results of characterizing the source over two
and a half days are shown in Fig. \ref{figure::source_stability}.
After
switching, the photon pairs are measured with a correlated photon detection
system (NuCrypt LLC, Model CPDS-4) consisting of 
an array of four InGaAs/InP avalanche
photodiodes.

\section{Switch Characterization}

In order to test the switch's effectiveness for quantum communications, we
measured both active and passive switching of both classical pulses and
polarization-entangled single-photon
pairs.  Figure \ref{figure::source-switch}(b) shows the switch as integrated
into the fiber-based source
of entangled photons referenced above.  To test multiple switching windows, loop
lengths of 500 m ($\approx$900-ps window) and 100 m ($\approx$180-ps window) were used. The
insertion loss introduced by  these switches in the O-band quantum channel
was
measured to be 1.3 dB ($L$ = 100 m, port T), 1.7 dB ($L$ = 100 m, port
R), 1.7 dB ($L$ = 500 m, port T), and 2.1 dB ($L$ = 500 m, port R).  Because
all of these directly measured losses include the 0.4 dB or 0.8 dB loss from one or
two passes through an optical
circulator, the raw switching loss for either transmission through or
reflection from the switching loop is 
0.9 dB (1.3 dB) for the $L$ = 100-m (500-m) loop.
%0.9 dB for the $L$ = 100-m loop or 1.3
%dB for the $L$ = 500-m loop.

\subsection{Switching contrast}

\begin{figure}
\centering
\includegraphics[width=3in]{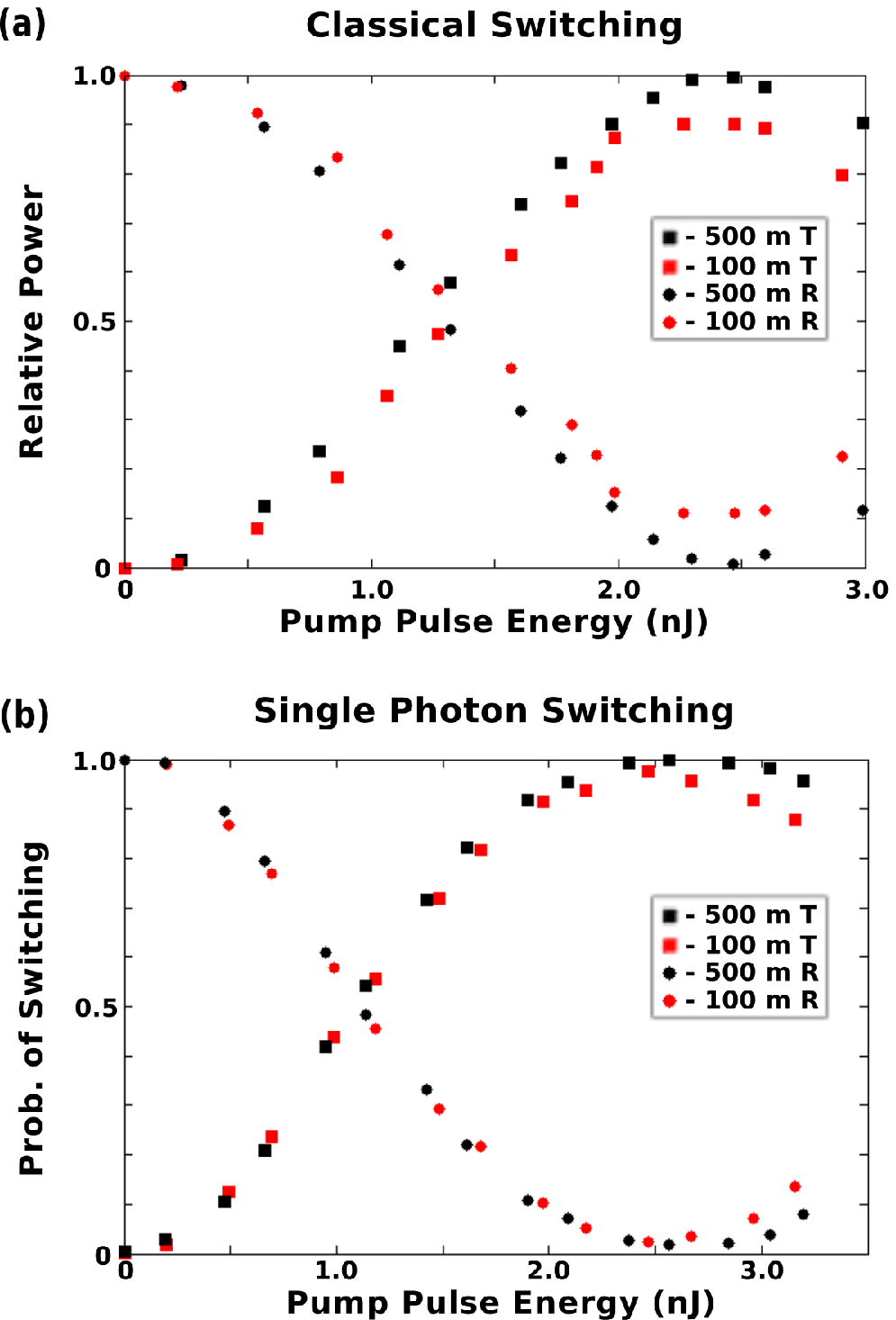}
\vspace{-1mm}
\caption{
(a) Classical switching contrast. Plot
shows normalized optical power at the switch outputs versus the energy of
the 1550-nm pump pulses.  
(b) Single-photon switching
contrast.  Plot shows the probability of routing an incoming single photon
versus pump energy for $L = 500$ m and $L = 100$ m (detector dark counts
subtracted).  
} \label{figure::contrast}
\vspace{-3mm}
\end{figure}

An important metric for both quantum and classical
routers is the switching \emph{contrast}---the ratio of
power directed to the desired output port divided by the
power directed to the complementary output port.  
Figure \ref{figure::contrast}(a) 
shows the classically measured switching contrast as a function
of the pump-pulse energy.  An optimal switching contrast of 150:1 is
achieved at a pump energy of 2.5 nJ/pulse for $L = 500$ m. Although the switching
contrast is expected to be independent of $L$, the 100-m data does not appear
to achieve full contrast---only 9.2:1.  This artifact is due to a long,
low-power tail ($\approx$370 ps total pulse width) in the 1305-nm test pulses
coupled with the slow response time of the optical detectors used to
collect this data. The 500-m switch opens an $\approx$1-ns wide switching
window---enough to fully switch the entire 370-ps classical pulse and thus
show full contrast.  The 100-m switch, however, opens a shorter, 200-ps
window and can fully switch only the portion of the energy contained in
that time window.  Repeating these tests using single photons that are
spontaneously generated from the same 1305-nm classical pulses
significantly reduces the effect of this artifact---compare Figs. 
\ref{figure::contrast}(a) and \ref{figure::contrast}(b).  Because
the photon-pair production rate is proportional to the pump-power squared,
a much smaller percentage of the generated photons resides in the tail of
the pulse. Consequently, the measured single-photon switching contrasts are
much closer to each other, 120:1 for the 500-m switch and 43:1 for the
100-m switch (see Fig. \ref{figure::contrast}(b)). We expect the true switching contrast to be the same in both
cases, because the tails, although quadratically reduced, still exist in
the single-photon test pulses used in these measurements.

\subsection{Temporal profile of the switching operation}

\begin{figure}
\centering
\includegraphics[width=3in]{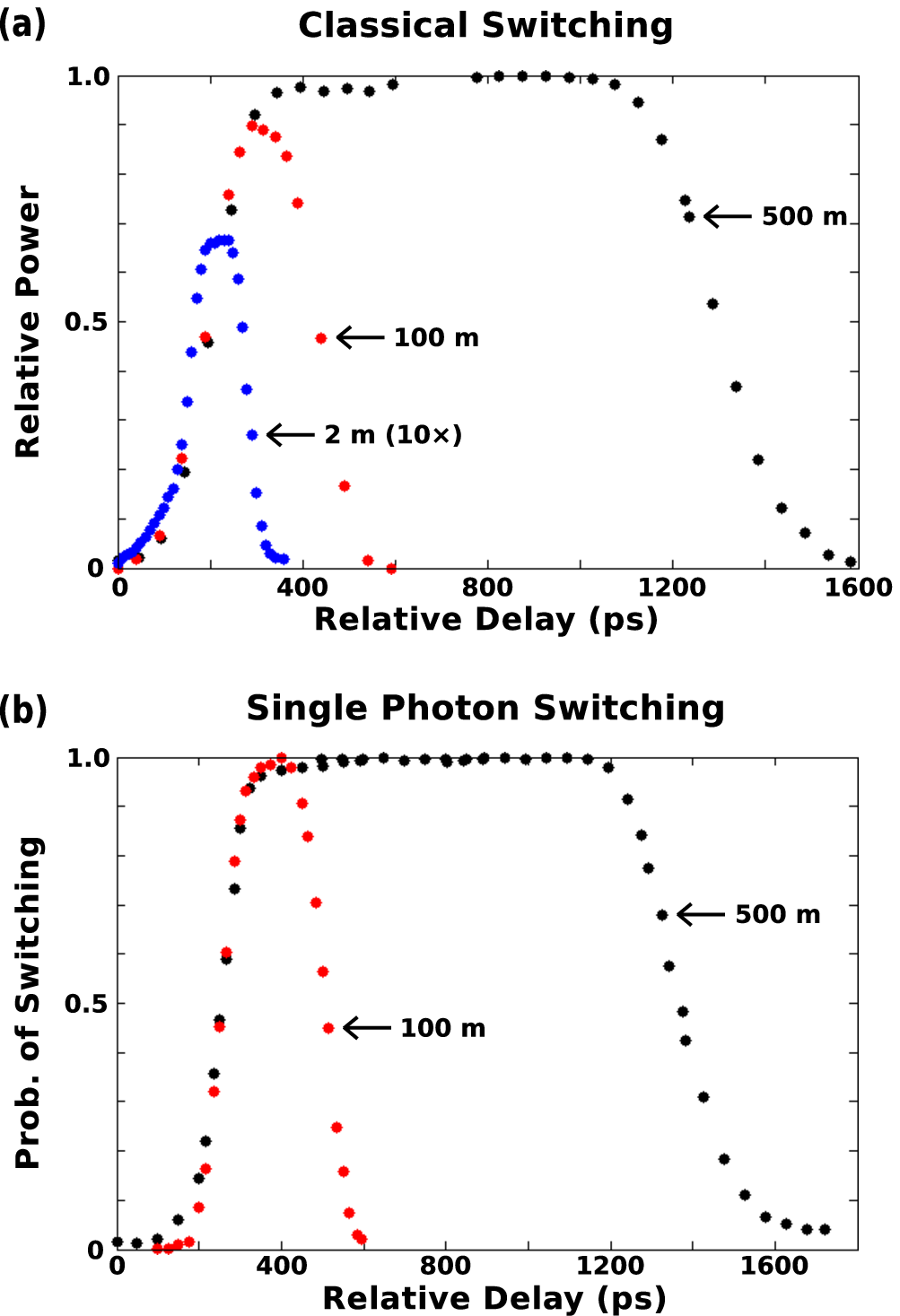}
\vspace{-1mm}
\caption{
(a) Temporal extent of the switching window, as
measured using classical test pulses. Plot shows optical power at the
switch output $T$ versus the relative delay between the pump and signal
pulses (2-m loop data is shown magnified $10\times$.).  
(b) Temporal extent of the switching window, as measured using
single photons.  Plot shows single-photon counts versus relative delay
between the single-photon and the pump pulses (detector dark counts
subtracted).
} \label{figure::window}
\vspace{-3mm}
\end{figure}

In addition to its use as a single-photon router, the switch is a
spatio-temporal coupler, enabling the encoding and decoding of quantum
information into a temporally multimode Hilbert space, which is, in
principle, boundless.  The extent to which this Hilbert space can be
effectively accessed, however, is determined by the temporal switching
profile of the device described above.  In order to characterize the
shape and width of the switching window, we introduce a relative delay
between the signal and the pump pulse paths.  

Although we are primarily
interested in the switch's single-photon performance, it is useful to first
characterize the switch's classical switching window.  Figure
\ref{figure::window}(a) shows the measured results for 500-m,
100-m, and 2-m loop lengths, which should correspond to 
$\approx850$ ps,   
$\approx170$ ps, and
$\approx4$ ps switching windows (full-width at half-maximum), respectively.  Surprisingly, each of the switching windows
shown in Fig. \ref{figure::window}(a) is $\approx150$--200 ps broader than these
estimates, even for the 2-m switch.  To understand this behavior, recall
that the real switching window must be the \emph{convolution} of the pump
pulses' temporal profile ($\approx5$ ps for our dual-color pump) with 
a square wave whose width is $\tau_s = 1.7$ ps per meter of
switching fiber.  The \emph{measured} switching window, however, must be
the convolution of the real switching window and
\emph{the temporal profile of the signal to be switched}.  Our classical,
1305-nm
test signal is in fact split from the same pump that is used in our entanglement
source, i.e., the 100--200-ps pulse described above.  In fact, the 2-m switching window shown in
Fig. \ref{figure::window}(a) gives us the most precise measure to date of
this pulse's temporal profile.

Because the probability of generating an entangled photon pair from this pump
pulse is proportional to the square of the pulse's instantaneous intensity,
we can expect the single-photon switching results to exhibit slightly
narrower---yet still broadened---temporal switching windows.   By sweeping
the relative delay while measuring
the switched single photons we map the switching window and observe this
effect (shown in Fig.
\ref{figure::window}(b)).  
As expected the temporal extent of the photons being switched blurs the true switching
window.  To quantitatively estimate the extent of this blurring, we
use the experimentally characterized 1305-nm classical pump pulses to
apply a numerical-fit deconvolution to the single-photon switching results.  In this way we obtain
the instantaneous temporal widths of the 100-m and 500-m switching windows
to be 180 ps and 900 ps, respectively, as verified by the single-photon test
results shown in Fig. \ref{figure::window}(b).

\subsection{Single-photon background}

\begin{figure}
\centering
\includegraphics[width=3.5in]{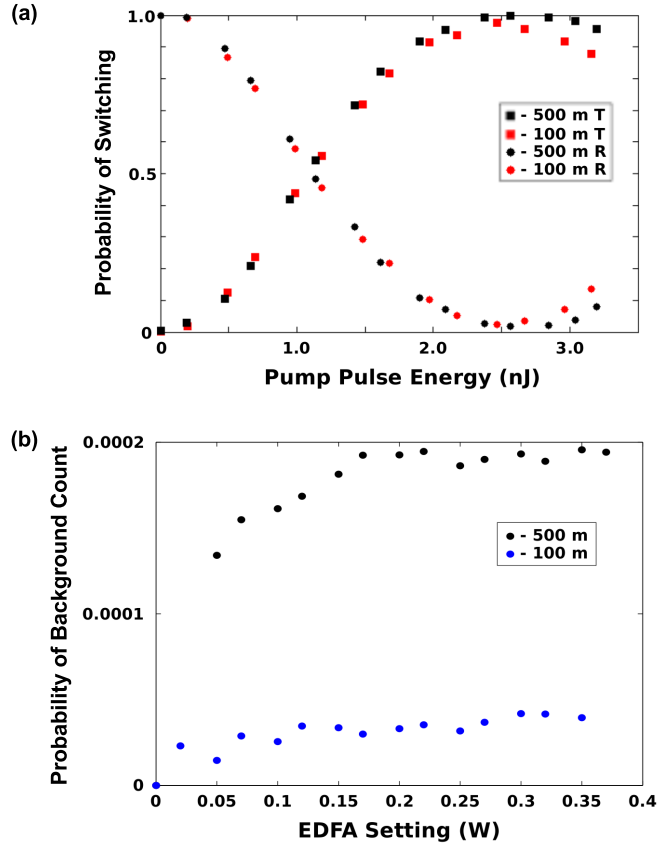}
\vspace{-1mm}
\caption{
(a) Switching probability as a function of pump pulse energy.
(b) Probability that a switching control pulse of a given energy will generate
a background photon count in the 1310-nm signal band.  
This probability is found to be proportional to $L$: 
$\approx 4 \times 10^{-7} \mathrm{m}^{-1}$.
We note that for the two data sets shown, the pump-pulse energies plotted
in (a) correspond to the EDFA settings in (b).
} \label{figure::background}
\vspace{-3mm}
\end{figure}

Closely related to contrast is the generated single-photon background,
from---for example---Raman scattering of the 1550-nm pump pulses.  Because
any in-band single-photon background has the potential to ``drown
out'' the entangled photons to be switched, they are a serious potential
source of error.  We
measured the 
probability of generating a 1310-nm background photon count and found it to
be proportional to $L$
($\approx 4 \times 10^{-7} \mathrm{m}^{-1}$).
Figure \ref{figure::background} shows the measured single-photon background
counts as a function of the pump power (measured at the pump EDFA's 
output, or ``EDFA setting'').  Because the background light
is found to be proportional to the fiber length, its genesis is consistent
with the Raman
scattering of the pump.  The negative effects of such scattering 
are expected to be much smaller for more advanced switches, because  
the chance that a background
photon would be created within the mode of interest (i.e., in the
same temporal mode as the pulse to be switched) may in practice be much
lower than measured here, and instead
be proportional to the temporal bandwidth: 
$\approx 2 \times 10^{-7} \mathrm{ps}^{-1}$.  
This low scattering probability is
consistent with the 35-THz Stokes-side detuning of the pump.  Although this
measured scattering probability is very low, the data is not fully
described by the Raman scattering alone; for example, the scattering probability
does not decrease linearly with EDFA output power but remains relatively
constant.  To get a clearer picture of the fundamental noise limitations of
this switch, more extensive testing with additional switch-loop lengths
will be required.

\subsection{Switching entangled single photons}

A crucial performance
benchmark for the switch is its ability to route entangled photons without
disturbing their quantum state.  In order to measure the extent to which
any disturbance occurs, we completely characterize the quantum state of
unswitched entangled photons with actively switched entangled photons.

\begin{figure}
\centering
\includegraphics[width=3.5in]{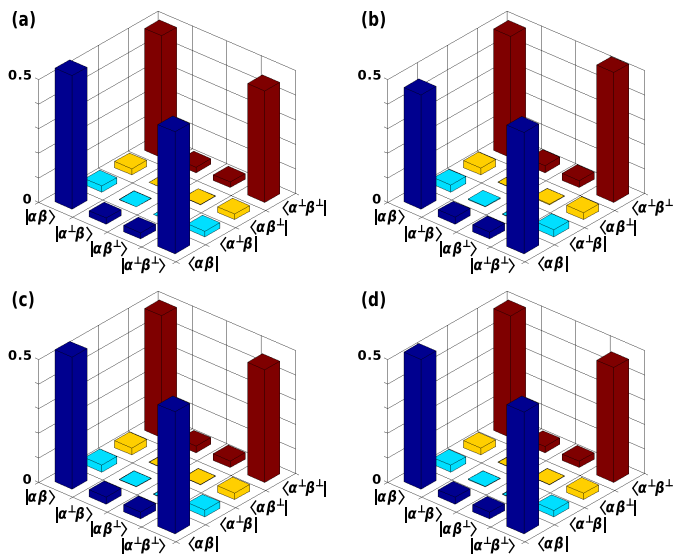}
\vspace{-1mm}
\caption{
(a) Reconstructed density matrix of the
unswitched state with the 500-m switch (fidelity to a maximally entangled
state, $F = 99.5\% \pm 0.2\%$; tangle, $T = 0.982 \pm 0.005$; linear
entropy, $S_L = 0.0005 \pm 0.003$). 
(b) Reconstructed density matrix of the switched state
from the 500-m switch ($F = 99.2\% \pm 0.2\%$, $T = 0.967 \pm 0.007$, 
$S_L = 0.01 \pm 0.005$). 
(c) Reconstructed density matrix of the unswitched state for 
$L$ = 100-m ($F = 99.5\% \pm
0.2\%$, $T = 0.981 \pm 0.01$, $S_L = 0.001
\pm 0.004$).  
(d) Density matrix of the switched state for $L$ = 100 m 
($F = 99.4\% \pm 0.4\%$, $T = 0.978 \pm 0.01$, $S_L = 0.005 \pm 0.01$).
} \label{figure::tomographies}
\vspace{-3mm}
\end{figure}

First, unswitched (no pump) entangled
photons from port R are characterized followed by actively switched
entangled photons from port T.  In both cases we use coincidence-basis
quantum-state tomography \cite{tomo1, tomo2}.  Both signal and idler
photons are analyzed
using separate quarter-waveplate (QWP), half-waveplate (HWP), and 
polarizing-beam-splitter (PBS) combinations, which together perform arbitrary single
qubit measurements. The measured coincidence rates---after subtracting
accidental coincidences, a procedure which lowers statistical errors
\cite{accidentals}---for 36
combinations of analyzer settings \cite{detecting_entanglement} are subjected to a numerical
maximum likelihood optimization, which reconstructs the density matrix most
likely to have produced the measured results.
Figures \ref{figure::tomographies}(a) and
\ref{figure::tomographies}(b), respectively, show the reconstructed density matrices for 
passively switched (port R) and actively switched (port T) entangled
photons, after reflection or transmission through the $L$ = 500-m loop.
Similar reconstructions for the $L$ = 100-m loop are shown in Figs.
\ref{figure::tomographies}(c) and
\ref{figure::tomographies}(d); 
in all four cases, the fidelity of the measured state
to a maximally-entangled state exceeds 99.0\%.  In addition, no measureable
state degradation results from active versus passive switching.

\subsection{Polarization dependence}

The quantum state tomography tests described above provide some
evidence that the switch is polarization independent (because a strong
polarization dependence would lead to a degradation in the polarization
entanglement).
To perform a more thorough study of the switch's polarization dependence, 
we introduce a 1-m ($\approx 5$ ns) delay into the
1545-nm arm of the grating filter used to generate the two-color
XPM pulses.  By reverting to a continuous wave (CW) O-band signal, the switching performance
for each color of the switching pulse can be independently monitored as
their polarizations are rotated using a single FPC.  Additionally,
adjusting the signal FPC does not 
change the switched output from either pump-color/polarization by more than
20\%.
This small change is attributed to the different birefringence-induced
polarization rotations experienced by
the O-band versus the C-band light.   In other words, even a single-polarization pump
pulse may have its polarization rotated relative to a given signal pulse,
resulting in relatively polarization-independent switching for a
single-color pump.  However, we anticipate that single-color-pump devices will
become more polarization dependent as the switching window is reduced
(i.e., as $L$ is decreased).

\begin{figure}
  \centering
  %\begin{tabular}{cc}
  %  \includegraphics[width=3in]{sepratecolorpowervedfa} &
  %  \includegraphics[width=3in]{sepratecolorphase} 
  %\end{tabular}
  \includegraphics[width=3in]{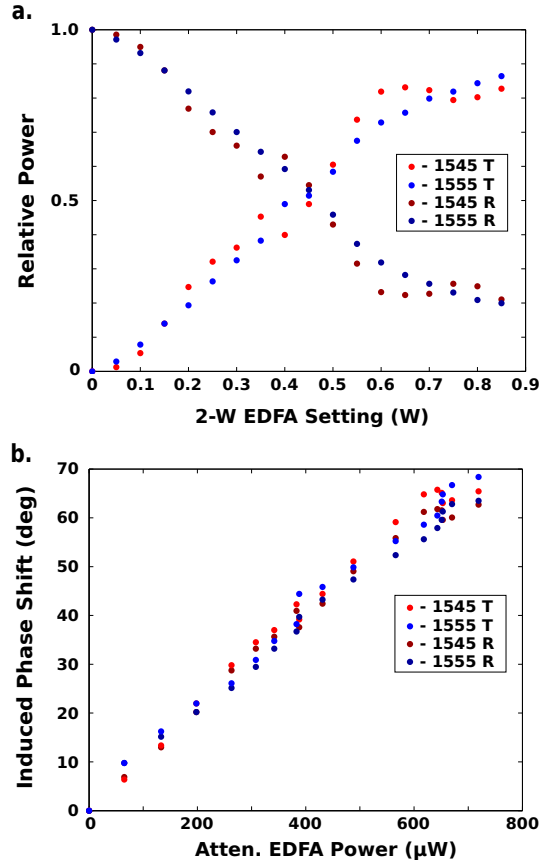} 
  \caption{
  Experimental results for the switch with the two colors of the
  XPM pump temporally separated. (a) Normalized power of the O-band signal reflected and
  transmitted by the temporally separated 1545-nm and 1555-nm
  XPM pump pulses versus the EDFA power settings. (b) Computed phase-shift
  applied by each color of the XPM pump for the measured EDFA powers.}
    \label{figure::sep_colors}
\end{figure}

Next, the two time-delayed pump powers are balanced and both output ports of the
switch are monitored on 1-GHz-bandwidth photodiodes connected to the same
oscilloscope, while the switch is initially set to reflect the entire incoming
CW signal.   In this way we can monitor the 
relative strength of the signal output pulses for
a range of EDFA settings;
the results are plotted in Fig. \ref{figure::sep_colors}(a). Measuring the relative
signal values at the reflected and transmitted ports allows us to calculate the
optical phase imparted by either pump color through the XPM process, as shown
in Fig. \ref{figure::sep_colors}(b).  The agreement between the calculated
values of the cross-phase shift from the transmitted and reflected ports
for both pump colors provides 
strong evidence that the results are interpreted correctly. 
It further shows that, when the two pump colors are recombined (by removing
the 1-m delay), the resulting pump pulses would act as the desired depolarized XPM
pump.  It also confirms that the EDFA provides a relatively constant gain for both
colors of the XPM pump.

\section{Demultiplexing a Dual-Channel Entangled Photon Stream}

The switch's ability to manipulate spatial and temporal quantum information has the
potential to enable new quantum communication protocols and new 
quantum networks. As an example of
this functionality, we use the switch to demultiplex a single quantum
channel from a dual-channel entangled photon stream.  The goal of this
experiment is to demonstrate a time-division-multiplexed (TDM) single-photon quantum channel
operating at 3-GHz.  To simulate this type of quantum network, we encode
two maximally entangled photon pairs into adjacent temporal modes,
separated by 300 ps.  After demonstrating that each of the adjacent temporal
modes contains a different maximally entangled state (see Figs.
\ref{figure::tdm}(a) and \ref{figure::tdm}(b)), we measure both
modes simultaneously, which results in a low-entanglement mixture of the two
different maximally entangled
states (see Fig. \ref{figure::tdm}(c)).  Finally, we use our single-photon switch to demultiplex the two
modes, recovering a single-channel's maximal entanglement (Fig.
\ref{figure::tdm}(d)).

\begin{figure}
\centering
\includegraphics[width=1.5in]{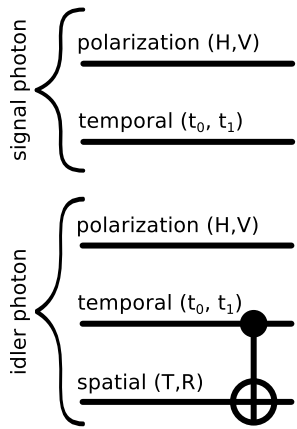}
\vspace{-1mm}
\caption{
Diagram of the five degrees of freedom in a multiplexed entangled
photon stream, which can be demultiplexed by applying the controlled
switch operation (shown).  
} \label{figure::encoding}
\vspace{-3mm}
\end{figure}

For this test, we encode the dual-channel entangled state into 
a five-qubit Hilbert space (see Fig. \ref{figure::encoding}) defined by the signal and idler polarization qubits
($\ket{H^{s,i}}, \ket{V^{s,i}}$),
the signal and idler temporal qubits
($\ket{t_0^{s,i}}, \ket{t_1^{s,i}}$), and an idler spatial qubit
($\ket{T^{i}}, \ket{R^{i}}$ (see Fig. \ref{figure::source-switch}(b)).
Using this encoding, we create the five-qubit hyper-entangled state 
\begin{equation}
    \ket{\Phi} = c_1 \ket{\psi_1} \ket{t_0^s} \ket{t_0^i} \ket{T^i} + 
                  c_2 \ket{\psi_2} \ket{t_1^s} \ket{t_1^i} \ket{T^i}, 
\end{equation}
where
\begin{eqnarray}
    \ket{\psi_1} &\equiv& \frac{1}{\sqrt{2}} \left( \ket{H^s}\ket{H^i} + \ket{V^s}\ket{V^i} \right),  \\
    \ket{\psi_2} &\equiv& \frac{1}{\sqrt{2}} \left( \ket{H^s}\ket{H^i} - \ket{V^s}\ket{V^i} \right), 
\end{eqnarray}
and $c_1$ and $c_2$
are arbitrary coefficients.  Measuring $\ket{\Phi}$ using
polarization-basis tomography while \emph{tracing out} temporal
degrees of freedom and \emph{projecting} into the idler spatial mode $\ket{T^i}$ 
will yield a highly mixed state, exactly the result
one expects from a simultaneous measurement of multiple entangled
quantum channels.  A switch capable of implementing a controlled-NOT
operation which couples the spatial and temporal qubits (see Fig.
\ref{figure::encoding}),  
however, would transform $\ket{\Phi}$ into the state:
\begin{equation}
\ket{\Phi^\prime} = c_1 \ket{\psi_1} \ket{t_0^s} \ket{t_0^i} \ket{T^i} + 
              c_2 \ket{\psi_2} \ket{t_1^s} \ket{t_1^i} \ket{R^i}.   
\end{equation}
This demultiplexed state should exhibit maximal
entanglement when \emph{projected} into the spatial mode $\ket{T^i}$, because even after
tracing over the temporal degrees of freedom only the maximally entangled
polarization state $\ket{\psi_1}$ would be present.

\subsection{Generating a dual-channel entangled photon stream}

\begin{figure}
\centering
\includegraphics[width=6in]{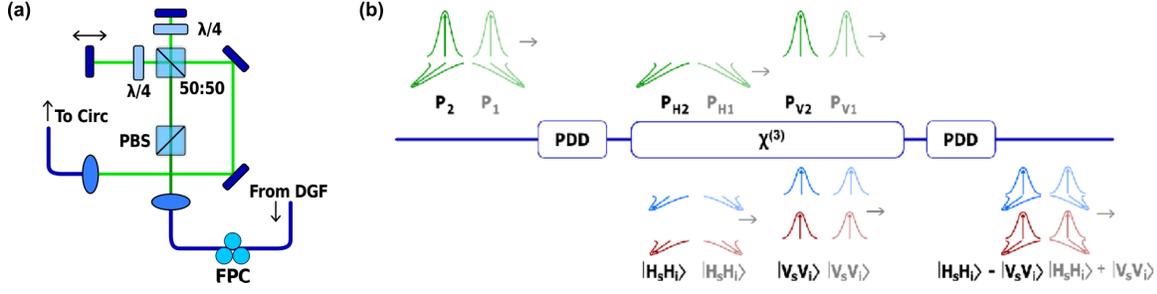}
\vspace{-1mm}
\caption{
(a) Michelson interferometer used for preparing the SFWM pump to generate 
time-division-multiplexed entangled photons.  (b) Pictorial diagram of the
SFWM processes which produce the time-division-multiplexed
entangled photon stream.
} \label{figure::demult}
\vspace{-3mm}
\end{figure}

To implement this proof-of-principle test of the switch's ability to couple
spatial and temporal degrees of freedom, 
we modify our
O-band entangled-photon source \cite{oband} by pumping it with a pair of 
pulses separated by 
$\Delta t \equiv t_1 - t_0 \approx 300 \mathrm{ps}$.  Fig.
\ref{figure::demult}(a) shows a diagram of the interferometer used to delay
one pump pulse relative to the other, while Fig. \ref{figure::demult}(b)
shows a pictorial view of the creation of the dual-channel entangled photon
stream.  Note that the same input pump pulse is split with a Michelson-type
interferometer with an additional quarter waveplate in each arm.  These
quarter waveplates
are aligned so as to minimize the light reflected back
through the initial PBS from one arm and to cut the reflected light
in half from the other.   Such settings lead to one vertically polarized pump pulse 
and one diagonally polarized
pump pulse temporally separated by $\approx 300$~ps.  
By choosing these polarizations for the leading and trailing pump pulses
(and after transmitting them through additional in-fiber polarization
rotations) the resulting 
pump state is
            $\sqrt{c_1} \left(\ket{H^p} +  \ket{V^p}\right) t_0 + 
             \sqrt{c_2} \left(\ket{H^p} + i\ket{V^p}\right) t_1$,
which upon SFWM gives the output signal-idler state $\ket{\Phi}$.
For the demultiplexing test, we choose
$c_1/c_2 \approx 1.25$ and $\Delta t \approx300$ ps.  
An experimental schematic for this type of pump preparation and
entangled photon generation is shown in Fig. \ref{figure::demult}.

\begin{figure}
\centering
\includegraphics[width=3in]{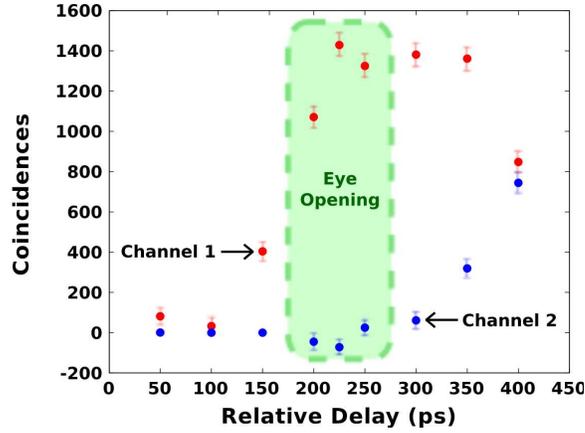}
\vspace{-1mm}
\caption{
Temporal profiles of the two independent channels in a dual-channel entangled photon
stream.  A single channel (either 1 or 2) is measured 
by blocking one arm of the Michelson interferometer
(see Fig. \ref{figure::demult}(a)) used to create the dual-channel
entangled photon source, in effect altering the source to create only a
single temporal channel of the entangled photon pairs.  The temporal profile of
a single entangled-photon channel is recorded using the $L=100$~m switching
device described above, configured to transmit any signal photons which
arrive during the 200-ps switching window defined by the C-band switching
pulse.  By recording the accidental-subtracted coincidence counts as a
function of the C-band pulse's arrival time (controlled using the global
TOD in the XPM pump path---see Fig. \ref{figure::dual_color_pump}), we can map the temporal profile of each
channel.
The red datapoints correspond to the state depicted in
Fig. \ref{figure::tdm}(a) while the blue datapoints correspond to the state
depicted in Fig. \ref{figure::tdm}(b).  By placing the global XPM-pump delay in the
central region marked with a green dashed box and unblocking both arms of
the source Michelson interferometer, we are able to demultiplex
channel 1 from channel 2.  The behavior
shown here is equivalent to a classical switching ``eye diagram''.
} \label{figure::eye}
\vspace{-3mm}
\end{figure}

Next, we determine the appropriate setting of the global delay
in the XPM pump path. We actively switch the idler channel to
port T and then measure the coincidences between the signal and
idler channels while blocking exactly one arm of the Michelson
interferometer shown in Fig. \ref{figure::demult}(a).  By recording 
coincidence counts for each configuration as a function of the global pump
delay, we are able to reconstruct two curves which are analagous to a
classical ``eye diagram'';
as shown in
Fig. \ref{figure::eye}.
Setting the delay to 225 ps, we achieve
full switching of the coincidences of the first temporal channel,
with minimal contribution from the second.

\subsection{Demultiplexing results}

\begin{figure}
\centering
\includegraphics[width=4in]{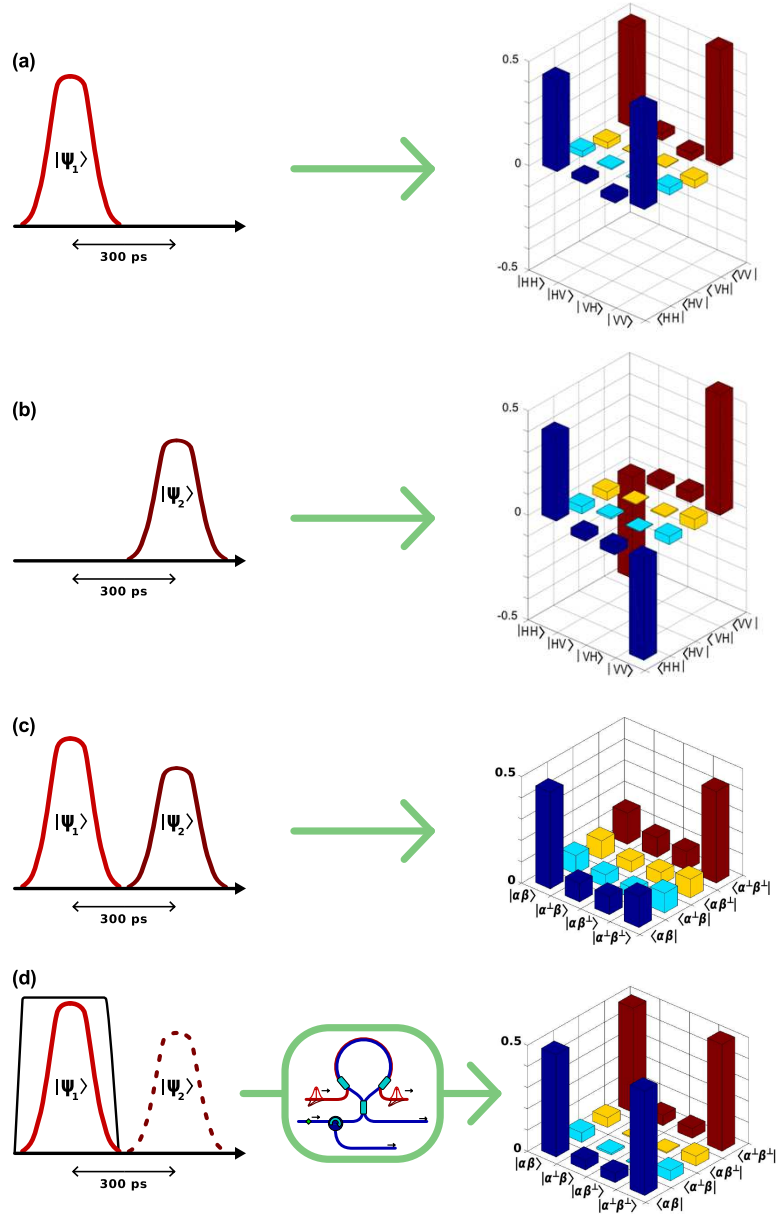}
\vspace{-1mm}
\caption{
Pictorial description of the construction and demultiplexing of a
dual-channel entangled photon stream.  (a) Channel 1, containing the
entangled state 
$\ket{\psi_1} \equiv \frac{1}{\sqrt{2}} \left( \ket{H^s}\ket{H^i} +
\ket{V^s}\ket{V^i} \right)$ (fidelity to a maximally entangled state $F >
99\%$).
(b) Channel 2, containing the entangled photon state
$\ket{\psi_2} \equiv \frac{1}{\sqrt{2}} \left( \ket{H^s}\ket{H^i} -
\ket{V^s}\ket{V^i} \right)$ ($F > 99\%$).
(c) Multiplexed channels,
defined by 
the five-qubit state
$\ket{\Phi} = c_1 \ket{\psi_1} \ket{t_0^s} \ket{t_0^i} \ket{T^i} + 
              c_2 \ket{\psi_2} \ket{t_1^s} \ket{t_1^i} \ket{T^i}$. 
By tracing over the temporal degree of freedom and projecting into the spatial
mode $\ket{T^i}$, we reconstruct a highly mixed density matrix with 
dramatically reduced state fidelity 
($F = 58.9\% \pm 0.5\%$).
(d) De-multiplexed channel.  By actively switching the channel 1 idler photons to
output T, c.f. Fig. \ref{figure::source-switch}(b) (i.e., demultiplexing channel 1), we can project into spatial mode
$\ket{T^i}$ and recover the
maximally entangled state $\ket{\psi_1}$ ($F = 98.6\% \pm 0.7\%$).
} \label{figure::tdm}
\vspace{-3mm}
\end{figure}

Figure \ref{figure::tdm}(c) shows
the experimentally measured density matrix for the multiplexed quantum
channels. As expected, the state is highly mixed; its fidelity to the
nearest maximally entangled state is only 58.9\%.  Utilizing the 100-m
switch we then
demultiplex (i.e., actively switch) only the first quantum channel 
($t = t_0$), creating the state $\ket{\Phi^\prime}$.  
As shown in Fig. \ref{figure::tdm}(d), after demultiplexing 
we are able to recover the high fidelity (98.6\%) of the target state
to a maximally entangled state.  
Because the cross-Kerr phase shift has previously been shown to
maintain spatial and temporal coherence in NOLM switches \cite{nolm1,
nolm2}, we anticipate that this switch's 
cross-Kerr-based demultiplexing operation is in fact coherent and equivalent to the
controlled-NOT operation depicted in Fig. \ref{figure::encoding} 
(although measurements equivalent to a full three-qubit process tomography would be necessary to show
this conclusively).
Moreover, unlike LOQC-based controlled-NOT gates, this switch is completely
deterministic and easily extensible, capable of independently tunable
couplings (e.g., 
controlled-$\pi/4$) to many temporal qubits encoded onto the same
photon  (by changing the control-pulse's
intensity as a function of time).  By cascading several switches, it is
also possible to couple to multiple spatial qubits.

\section{Conclusion}

We have demonstrated the first all-optical 
switch suitable for single-photon quantum communications. It achieves
low-loss ($< 1$ dB when used to switch between transmitted and reflected
modes, $< 1.7$ dB when combined with a circulator), 
high-isolation ($> 20$ dB), and high-speed ($< 200$ ps) performance without
a measureable disturbance to the quantum state of the routed single
photons. We demonstrate its ultrafast capability by demultiplexing a single
quantum channel from a time-division-multiplexed stream of entangled
photons.  Very few fundamental limitations apply to this type
of switch design.  With carefully designed fiber components, one has the
potential to dramatically reduce the switch's loss.  In principle the only
unavoidable switching losses are fiber transmission losses (0.15--0.2 dB/km) and circulator
insertion losses (waveguide-based circulators with a
0.05 dB insertion loss have been designed and simulated \cite{waveguide-circulators}).
Additionally, decreasing
$L$ to a few meters will reduce the switch's speed to
$\approx$10 ps
while simultaneously decreasing the background by an order of magnitude.  Even without
these improvements, however, this switch represents an important new tool
for manipulating spatianlly- and temporally-encoded quantum information.

This research was supported in part by the DARPA ZOE program (Grant No.
W31P4Q-09-1-0014) and the NSF IGERT Fellowship (Grant No. DGE-0801685).

%\section{Methods}

%\subsection{Pump preparation for the entangled photon source}
%100-ps pulses at 50-MHz repetition rate were carved into the output of a CW
%laser (Santec, Model TSL-210V) using a 10-GHz bandwidth amplitude modulator
%(EO Space, Model AZ-OK5-10) clocked off of the XMP-pump laser. In order to
%prevent the extinction ratio of the modulator from drifting during data
%collection, the bias voltage was modulated by a weak 100.5 kHz electrical
%signal. Light from the 10 percent output of a 90:10 fiber coupler after the
%PDFA (Fiberlabs, Model AMP-FL8611-OB) was detected and fed to a lock-in
%amplifier, which provided a continuous bias correction feedback to the
%modulator.


\section*{References}
\begin{thebibliography}{}
\vspace{-4mm}
%666

\bibitem{mike_and_ike} M. A. Nielsen and I. L. Chuang, \emph{Quantum Computation
            and Quantum Information} (Cambridge Univ. Pr., 2000).
\bibitem{nweke} N. I. Nweke, \emph{et al.} 
            \emph{Appl.  Phys. Lett.} \textbf{87}, 174103 (2005).
\bibitem{oband} M. A. Hall, J. B. Altepeter, and P. Kumar, 
            \emph{Optics Express} \textbf{17}, 14558 (2009).
\bibitem{previous-first} M. A. Duguay and J. W. Hansen, 
            \emph{Appl. Phys. Lett.} \textbf{15}, 192 (1969).
\bibitem{previous2} N. J. Doran, and D. Wood, 
            \emph{Opt. Lett.} \textbf{13}, 56--58 (1988).
\bibitem{mems1} K. Hogari and T. Matsumoto,
            \emph{Appl. Opt.} \textbf{30}, 1253--1257 (1991).
\bibitem{nolm1} M. Eiselt, 
            \emph{Electron Lett.} \textbf{28}, 1505 (1992).
\bibitem{previous5} J. P. Sokoloff, P. R. Prucnal, I. Glesk, M. Kane, 
            \emph{IEEE Photon. Technol. Lett.} \textbf{5}, 787 (1993).
\bibitem{previous6} M. Asobe, I. Yokohama, H. Itoh, and T. Kaino, 
            \emph{Opt. Lett.} \textbf{22}, 274 (1997).
\bibitem{previous7} I. Yokohama \emph{et al.}, 
            \emph{J. Opt. Soc. Am. B} \textbf{14}, 3368 (1997).
\bibitem{previous8} G. S. Kanter, P. Kumar, K. R. Parameswaran, and M. M. Fejer, 
            \emph{IEEE Photon. Technol. Lett.} \textbf{13}, 341 (2001).
\bibitem{previous9} J. E. Sharping, M. Fiorentino, P. Kumar, and R. S. Windeler, 
            \emph{IEEE Photon. Technol. Lett.} \textbf{14}, 77 (2002).
\bibitem{previous10} V. Van, \emph{et al.} 
            \emph{IEEE Photon. Technol. Lett.} \textbf{14}, 74 (2002).
\bibitem{waveguide_res1} V. R. Almeida \emph{et al.}, 
            \emph{Opt. Lett.} \textbf{29}, 2867--2869 (2004).
\bibitem{previous-last} G. Bertocchi, \emph{et al.} 
            \emph{J. Phys. B} \textbf{39} 1011 (2006).
\bibitem{eospace} http://www.eospace.com
\bibitem{waveguide_res2} P. Dong, S. F. Preble, and M. Lipson, 
            \emph{Opt. Express} \textbf{15}, 9600--9605 (2007).
\bibitem{mems2} C. Knoernschild, C. Kim, F. P. Lu, and J. Kim,
            \emph{Opt. Express} \textbf{17}, 7233--7244 (2009).
%\bibitem{acousto-optic} RECENT TEXTBOOK CITE.
\bibitem{nolm2} K. Uchiyama \emph{et al.},
            \emph{J. Lightw. Tech.} \textbf{15}, 194--201 (1997).
\bibitem{cband_noise} Q. Lin, F. Yaman, and G. P. Agrawal, 
            \emph{Phys. Rev. A} \textbf{75}, 023803 (2007).
\bibitem{sintec} http://www.sintecoptronics.com/
\bibitem{polarization_based} S. Kinoshita et al., 
            \emph{Rev. Sci. Instrum.} \textbf{71}, 3317 (2000).
\bibitem{waveguide_thermal} I. Kiyat, A. Aydinli and N. Dagli,
            \emph{Photonics Technology Letters}, IEEE, \textbf{18} 364--366 (2006).
%\bibitem{relative_performance} need long explanation here explaining our
    %methodology for being orders of magnitude better than all other cited
    %switches.
\bibitem{xpm} H. B\"{u}low and G. Veith, 
            \emph{Elect. Lett.} \textbf{29}, 588--589 (1993).
Note that if one of the two pump colors leads the other by a
time $\delta$, then the leading and trailing $\delta$-length segments of the
switching window will not be polarization-independent.
\bibitem{dual-lambda} K. J. Blow, N. J. Doran, B. K. Nayar, and B. P. Nelson, 
            \emph{Opt. Lett.} \textbf{15}, 248--250 (1990).
\bibitem{loop-mirror} D. Mortimer, 
            \emph{J. Lightw. Tech.} \textbf{6}, 1217--12124 (1989).
\bibitem{tomo1} D. F. V. James, P. G. Kwiat, W. J. Munro, and A. G. White,
            \emph{Phys. Rev. A} \textbf{64}, 052312 (2001).
\bibitem{tomo2} J. B. Altepeter, E. R. Jeffrey, and P. G. Kwiat, 
            \emph{Advances in AMO Physics, Vol. 52}, Ch. 3 (Elsevier, 2006).
\bibitem{accidentals} 
Increasing this source's pair production rate (PPR)
and then subtracting accidental coincidences increases measurement
precision while accurately predicting the low-PPR, non-accidental
subtracted result \cite{oband}.  Without this correction, high-PPR,
dark-count-only subtracted fidelities for the data shown in Figs. 1 and 3 are between
80--95\%.
\bibitem{detecting_entanglement} J. B. Altepeter, \emph{et al.}
            \emph{Phys. Rev. Lett.} \textbf{95}, 033601 (2005).
\bibitem{waveguide-circulators} R. Takei and T. Mizumoto,
\emph{Jpn. J. Appl. Phys.} \textbf{49}, 052203 (2010).

\end{thebibliography}
\end{document}